\documentclass[reprint]{revtex4}
\usepackage{amstext}
\usepackage{amsmath}
\usepackage{graphicx}
\usepackage{endnotes}
\usepackage{amssymb}

\def\be{\begin{equation}}
\def\ee{\end{equation}}
\def\bea{\begin{eqnarray}}
\def\eea{\end{eqnarray}}

\begin{document}

\title{Circular orbits on a warped spandex fabric}

\author{Chad A. Middleton}
\email{chmiddle@coloradomesa.edu}
\affiliation{Department of Physical and Environmental Sciences, Colorado Mesa University, Grand Junction, CO 81501, U.S.A.}

\author{Michael Langston}
\email{mlangsto@mavs.coloradomesa.edu}
\affiliation{Department of Physical and Environmental Sciences, Colorado Mesa University, Grand Junction, CO 81501, U.S.A.}

\begin{abstract}
We present a theoretical and experimental analysis of circular-like orbits made by a marble rolling on a warped spandex fabric.   We show that the mass of the  fabric interior to the orbital path influences the motion of the marble in a nontrivial way, and can even dominate the orbital characteristics.   We also compare a Kepler-like expression for such orbits to similar expressions for orbits about a spherically-symmetric massive object in the presence of a constant vacuum energy, as described by general relativity.
\end{abstract}\maketitle

\section{Introduction}

In Einstein's theory of general relativity (GR), gravity is described as the warping of space and time due to the presence of matter and energy.  In GR gravity is understood not as a force of one massive object acting on another, but rather as the manifestation of an object residing in the warped spacetime of another.  In Newtonian gravitation planetary orbits arise as the result of a planet encountering the attractive force of the Sun. In GR, orbits arise from geodesics of free bodies moving in curved spacetime, in which their ``straight line'' paths are curved due to the non-Euclidean geometry involved. 

In a typical introductory GR course, the spacetime geometry external to a spherically-symmetric, non-rotating massive object, which corresponds to the Schwarzschild solution, is studied extensively as it offers one of the simplest, realistic, and exact solutions to the field equations of GR.\cite{hartle, carroll}  To further understand the non-Euclidean nature of this Schwarzschild geometry, an embedding diagram is often constructed.  The procedure for generating such a diagram is as follows. The full four-dimensional geometry is examined at a moment in time, with the spherical polar angle set to $\pi/2$.  This amounts to examining a two-dimensional equatorial ``slice'' of the three-dimensional space.  The metric that describes the geometry of this two-dimensional slice can be equated to the metric of a curved two-dimensional surface residing (or embedded) in a flat three-dimensional space. Hence, one obtains a curved two-dimensional surface with the \textit{same} spatial curvature as the two-dimensional spatial slice of the original Schwarzschild geometry.   Because we can visualize  a curved two dimensional surface embedded in a flat three dimensional space, such diagrams prove to be quite useful in understanding the notion of curvature in GR and are often found on the covers of general relativity texts.~\cite{hartle}

When geodesics (or free particle orbits) are then considered in the Schwarzschild geometry, a conceptual analogy is often drawn to a small spherical object, such as a marble, rolling freely on an elastic fabric that is warped by the presence of a massive object placed upon its surface.  Presumably this warped  fabric resides in the constant gravitational field of the Earth with a central object pulled downward, its presence of which warps the fabric.  This analogy between geodesics in the Schwarzschild geometry and the orbits of a marble on a warped surface is fundamentally different than the connection between the embedding diagrams of a curved two-dimensional surface and the two-dimensional spatial slice of the Schwarzschild geometry.   The function that describes the shape of the embedded surface of the Schwarzschild geometry is relatively simple.   In contrast, there does not exist a two-dimensional, cylindrically-symmetric surface that will yield rolling marble orbits that are equivalent to the particle orbits of Newtonian gravitation,\cite{English} or for particle orbits that arise in general relativity.  Nonetheless, it is hard to argue that the conceptual analogy of a marble rolling on a warped fabric does not benefit the beginning student of general relativity.

White and Walker first explored the circular-like orbits of a marble on a warped spandex fabric where they employed a Newtonian treatment in determining the shape of the fabric and restricted their analysis to the small curvature regime.\cite{GW}    Interestingly, they found a Kepler-like expression of the form $T^3\propto r^2$, which is reminiscent of Kepler's third law for planetary orbits, but with the powers transposed.  Later, Lemons and Lipscombe showed that the shape of spandex fabric in the small curvature regime can be arrived at by employing a Lagrangian treatment that minimizes the total potential energy of the spandex fabric.\cite{DL}  The Lagrangian approach proves advantageous, as the results of White and Walker arise as a special case of a more general treatment where the pre-stretching of the spandex fabric can be included in the analysis.  The contributions to the total potential energy of the fabric considered in the work of Lemons and Lipscombe include the gravitational potential energy of the central mass and the elastic potential energy of the fabric.  

Here, we're interested in generalizing the analysis of circular orbits of a marble on a warped fabric in two ways.  First, in determining the shape of  the fabric we also employ a Lagrangian treatment of minimizing the total potential energy of the fabric including its gravitational potential energy.  This contribution was previously ignored and is shown here to influence the orbits of a marble in a nontrivial way.  Second, we offer an analysis of the shape of spandex fabric in both the small and large curvature regimes.  We find excellent agreement between the theoretical framework and the experimental results in both regimes \textit{only} after it is realized that the modulus of elasticity is not a constant for spandex fabric.  Last, we present  a thorough analysis of the circular-like orbits that arise in both the small curvature regime (where the warping of the fabric is small) and in the large curvature regime (where the warping is large).

Our paper is outlined as follows.  In Sec.~\ref{eqn} we consider the motion of a marble rolling on a generic, cylindrically-symmetric surface.  By constructing the Lagrangian in cylindrical coordinates and subjecting it to a constraint---the marble must reside on top of the surface---we obtain an equation of motion that describes the dynamics of the rolling marble.  In Sec.~\ref{Shape} we arrive at a non-linear, ordinary differential equation that dictates the shape of an elastic fabric by minimizing its total potential energy.  This potential energy includes the elastic potential energy of the fabric as well as the gravitational potential energy of the central mass \textit{and} that of the fabric.  In Sec.~\ref{orbits} we explore the circular-like orbits in the small curvature regime.  A Kepler-like relation connecting the radius of the orbit, the period of the motion, and the mass is found.  This Kepler-like relation reduces to that of White and Walker's when the gravitational potential energy contribution of the fabric is set to zero.\cite{GW}  It is shown, by measuring the radius  and the period of the rolling marble's orbit for a wide variety of central masses, that the gravitational potential energy term for the  fabric can dominate over that of the central mass.  In Sec.~\ref{orbits2} we explore the circular-like orbits in the large curvature regime.   When the marble is orbiting deep inside the well we again measure the period and the radius of orbit.  Using the measured values of the modulus of elasticity for large central masses, the radius of the orbit, and the period of revolution, we find good agreement with the theoretical model.  Lastly, in the Appendix~\ref{AdS}, we briefly outline the general relativistic treatment for obtaining circular orbits about a spherically-symmetric massive object in the presence of a constant vacuum energy.  After arriving at an exact Kepler-like relation for this orbiting body, we compare it to the approximate Kepler-like relation of the marble on the warped  fabric in the small curvature regime.

\section{The Equation of Motion for a Marble rolling on a cylindrically-symmetric surface} \label{eqn}

We begin by constructing the Lagrangian in cylindrical coordinates for a rolling marble, which is of the form
\be\label{L}
L=\frac{1}{2}m\left[\dot r^2+r^2\dot{\phi}^2+\dot{z}^2\right]+\frac{1}{2}I\omega^2-mgz,
\ee
where a dot indicates a time-derivative and $m$ is the mass of the marble. In addition to the translational and rotational kinetic energy terms, we have included the gravitational potential energy of the marble at height $z$.  The rotational kinetic energy in Eq.~\eqref{L} becomes $mv^2/5$, where the marble with radius $R$ and inertia $I=2mR^2/5$ rolls without slipping.

As here we are interested in examining the motion of a marble rolling on a warped fabric, there exists an equation of constraint connecting two of our coordinates, which is generically of the form $z=z(r)$.  The functional relationship connecting these two coordinates will be explicitly arrived at in the next section.   Thus, Eq.~\eqref{L} becomes
\be\label{Lagrange}
L=\frac{7}{10}m\left\{\left[1+z'(r)^2\right]\dot r^2+r^2\dot{\phi}^2\right\}-mgz(r),
\ee
where $z'(r)\equiv dz(r)/dr$.  Notice that we employed the chain rule when calculating time derivatives of $z(r)$, as shown elsewhere.~\cite{English}

Upon examination of Eq.~\eqref{Lagrange}, we note that the Lagrangian contains two generalized coordinates and will therefore yield two Lagrange equations of motion.  However, as the Lagrangian does not explicitly depend on the coordinate $\phi$,  the corresponding Lagrange equation will yield conservation of angular momentum about the $z$-axis, as $\phi$ represents a cyclic coordinate.\cite{Marion}  The relevant equation of motion that comes from considering the Lagrange equation for the $r$ coordinate is 
\be\label{rcoord}
\frac{\partial L}{\partial r}-\frac{d}{dt}\left(\frac{\partial L}{\partial\dot{r}}\right)=0,
\ee
and inserting Eq.~\eqref{Lagrange} into \eqref{rcoord} yields the dynamical equation of motion for the marble:
\be\label{EoM}
(1+{z'}^2)\,\ddot r+z'z''\,\dot{r}^2-r\dot{\phi}^2+\frac{5}{7}\,gz'=0.
\ee
As was previously shown,~\cite{English} Eq.~\eqref{EoM} will \textit{not} yield the Newtonian-like orbits of planetary motion for a marble on \textit{any} cylindrically-symmetric surface, except in the special case of circular orbits.  This discrepancy for noncircular orbits is due to the $\dot{r}^2$ term in Eq.~\eqref{EoM}, which is absent from the radial equation of motion in Newtonian theory.  It arises from motion in the $z$-direction when $\dot{r}\neq 0$.   It is noted that this inconsistency holds when considering a fully general relativistic treatment of planetary motion around a spherically-symmetric, non-rotating central object.   When considering geodesics about a spherically-symmetric, non-rotating central object, an additional $1/r^3$ term is added to the effective potential, in addition to the $1/r$ Newtonian and the $1/r^2$ angular momentum terms (see Appendix~\ref{AdS}).  This additional GR term does offer a small-$r$ modification to non-relativistic Newtonian orbits, but doesn't allow for consistency with Eq.~\eqref{EoM} for any cylindrically-symmetric surface described by $z(r)$.

In this work we simplify our analysis by focusing on circular orbits of a rolling marble on an elastic surface, where $\dot{r}=\ddot{r}=0$.  For this special case Eq.~\eqref{EoM} reduces to the concise form
\be\label{cirEoM}
\frac{4\pi^2r}{T^2}=\frac{5gz'(r)}{7},
\ee
where $T$ is the period of revolution and we used the fact that 
\be
v=r\dot{\phi}=\frac{2\pi r}{T}
\ee
for circular orbits.  Upon examination of Eq.~\eqref{cirEoM} one finds the exact Kepler-like relationship describing the circular orbits of the marble depends linearly on $z'(r)$, which equates physically to the slope of the elastic surface.  In the next section we minimize the total potential energy to arrive at a nonlinear ordinary differential equation that determines the shape of the warped elastic fabric.  An exact analytical expression for the shape of the fabric cannot be determined, but one can find the approximate behavior in the small and large curvature regimes, when the marble is orbiting far from the central mass and deep within the well.

\section{The shape of an elastic fabric warped by mass}\label{Shape}

We now wish to determine the shape of a 2D elastic fabric warped by a central mass placed upon the surface of the fabric.  The functional shape of this fabric is uniquely determined by the ``springiness'' of the fabric and by the weight of the central mass placed upon its surface.  When the mass of the central object is small, as compared to the mass of fabric itself, one must account for the mass of the fabric to accurately model this system.   To obtain the governing equation that determines the shape of the fabric, we first construct the total potential energy of the elastic fabric-mass system and then apply the principle of least potential energy---the fabric-mass system will assume the shape that minimizes the total potential energy of the system.

Consider a circular elastic fabric of radius $R$, which is initially unstretched and then subjected to a central mass placed upon its surface.   The elastic potential energy stored in a differential segment of the fabric, namely a concentric ring of radius $r$ and unstretched width $dr$, is of the form
\be\label{diffU}
dU_{e}=\frac{1}{2}\kappa(\sqrt{dr^2+dz^2}-dr)^2,
\ee
where $\kappa$ is the spring constant of the thin ring, $dr$ is the unstretched differential width of the concentric ring, and $\sqrt{dr^2+dz^2}$ is the stretched differential width.   The spring constant of the differential segment is in fact a function of the unstretched length and the radius of the concentric ring being considered, and can be related to the modulus of elasticity $E$ by
\be\label{E}
E=\frac{\kappa\;dr}{2\pi r},
\ee
where $E$ is a property of the elastic material and is assumed to be constant for this elastic surface.\cite{DLbook}  Equation~\eqref{E} can be understood conceptually as follows.  For an elastic ring of radius $r$ and variable width $dr$, a ring with twice the width will have a spring constant of half its value.  For an elastic ring of width $dr$ and variable radius $r$, a ring with twice the radius will have a spring constant of twice its value.  Thinking of each elastic ring as consisting of tiny massless springs, \cite{GW} a ring with twice the radius will have twice the circumference and therefore twice the number of springs in parallel, and twice the spring constant.

Employing Eq.~\eqref{E} and integrating Eq.~\eqref{diffU} over the whole elastic fabric of radius $R$, we obtain the total elastic potential energy of the fabric to be
\be\label{elaspot}
U_e=\int_0^R \pi E r\left(\sqrt{1+z'(r)^2}-1\right)^2\;dr.
\ee
This relation for the elastic potential energy of an elastic fabric was previously found,~\cite{DL}  where the stress applied to the edge of an annular ring of fabric was related to its strain.

We now calculate the gravitational potential energy of the fabric.  For an initially unstretched fabric allowed to hang freely, fixed only at the perimeter, the gravitational potential energy of a differential segment of the  massive fabric is given by the expression
\be\label{gravpot}
dU_{g,s}= dm_s \,g z(r),
\ee
where $z(r)$ is the height of the differential mass $dm_s$.  Notice that a fabric of uniform areal mass density will deviate from uniformity as the fabric stretches due to its own weight.   As the differential mass of a concentric ring is conserved, it obeys the relation
\be\label{sigma}
dm_s=\sigma_0 2\pi r \,dr=\sigma(z') 2\pi r \sqrt{dr^2+dz^2},
\ee
where $\sigma_0$ is the initial uniform areal mass density and $\sigma(z')$ is the variable areal mass density of the fabric as it hangs freely and is stretched.\cite{DLbook}  Using Eq.~\eqref{sigma} and integrating Eq.~\eqref{gravpot} over the whole fabric we obtain the total gravitational potential energy to be 
\be\label{gravpot2}
U_{g,s}=\int_0^R 2\pi\sigma_0g rz(r)\,dr.
\ee
Lastly, we need to construct the gravitational potential energy of the central mass in terms of an integral, which will allow for the three potential energy terms to be combined.  Thus, the gravitational potential energy of the central mass can be written as
\be\label{pointpot}
U_{g,M}=Mg z(0)=-\int_0^R Mg z'(r)\,dr,
\ee
where $M$ is the mass of the central object, which we approximated as being point-like.\cite{DL} Here, the gravitational potential energy of the fabric [Eq.~\eqref{gravpot2}] is also considered when constructing the total potential energy of the fabric-mass system. This contribution will be shown to dominate over Eq.~\eqref{pointpot} for a small enough central mass. 

The total potential energy of the fabric-mass system can be written in terms of a single integral of the form
\be\label{total}
U=U_e+U_{g,s}+U_{g,M}=\int_0^R f\left(z(r),z'(r);\,r\right)\,dr,
\ee
where we define the function $f$ to be
\be\label{functional}
f\left(z(r),z'(r);\,r\right)\equiv \pi E r\left(\sqrt{1+z'(r)^2}-1\right)^2+2\pi\sigma_0g rz(r)-Mg z'(r).
\ee
Here we have included the elastic potential energy of the fabric, the gravitational potential energy of the fabric of uniform areal density $\sigma_0$, and the gravitational potential energy of the central object of mass $M$. 

To minimize the total potential energy we subject Eq.~\eqref{functional} to the Euler-Lagrange equation
\be
\frac{\partial f}{\partial z}-\frac{d}{dr}\left(\frac{\partial f}{\partial z'}\right)=0,
\ee 
which yields a differential equation that determines the shape of the fabric warped by the mass.  This ordinary differential equation takes the form
\be
\frac{d}{dr}\left[rz'\left(1-\frac{1}{\sqrt{1+z'^2}}\right)-\frac{Mg}{2\pi E}\right]= \frac{\sigma_0g}{E} r,
 \ee
where we withheld writing the functional dependence of $z(r)$ for the sake of brevity.  This equation can be integrated and yields the expression 
\be\label{shape}
rz'\left(1-\frac{1}{\sqrt{1+z'^2}}\right)=\alpha\left(M+\pi\sigma_0 r^2\right),
\ee
where we set the constant of integration equal to zero and define the parameter
\be\label{alpha}
\alpha\equiv\frac{g}{2\pi E}.
\ee
Notice that the parameter $\alpha$ is inversely proportional to the modulus of elasticity $E$ and has units of m/kg.  Also notice that when one sets the areal mass density of the fabric equal to zero, Eq.~\eqref{shape} reduces a result previously presented.\cite{DL}  Equation~\eqref{shape} is a nonlinear, ordinary differential equation that cannot be solved analytically.  In the next sections we explore the behavior of this equation in the small and large $z'$ regimes.  Physically, this corresponds to the approximate behavior of the slope of the elastic fabric in the nearly flat (small curvature) regime and deep inside the well (large curvature).

\section{Circular orbits in the small curvature regime}\label{orbits}

We first explore the behavior of Eq.~\eqref{shape} in the nearly flat regime of the spandex fabric, where the warping of the fabric is small, namely, when $z'(r)\ll 1$.  In this small $z'$ regime we expand the square root in Eq.~\eqref{shape} in powers of $z'$ and keep only the lowest-order non-vanishing term to obtain
\be
z'(r)\simeq \left(\frac{2\alpha}{r}\right)^{1/3}\left(M+\pi \sigma_0 r^2\right)^{1/3}.
\ee 
Plugging this expression into Eq.~\eqref{cirEoM} and rearranging, we obtain the Kepler-like relation
\be\label{kepler}
T^3=\left(\frac{28\pi^2}{5g}\right)^{3/2}\frac{1}{\sqrt{2\alpha}}\frac{r^2}{\left(M+\pi \sigma_0 r^2\right)^{1/2}}.
\ee
Notice if we take $M\gg\pi\sigma_0r^2$---equivalent to neglecting the gravitational potential energy of the elastic fabric---Eq.~\eqref{kepler} reduces to
\be\label{white}
T^3\propto r^2/\sqrt{M}.
\ee
This equation corresponds to the result found previously,~\cite{GW} where a Newtonian treatment of the fabric was used in determining its shape.  Equation~\eqref{white} was later found~\cite{DL} by applying the principle of least potential energy to the fabric, similar to the approach presented here.\cite{noprestretch}  However, the aforementioned analyses neglected the gravitational potential energy of the fabric, which is relevant when the mass of the central object is small compared to the total mass of the fabric interior to the orbiting marble.

Equation~\eqref{kepler} relates the period of revolution $T$ to the radius of the orbit $r$ for a small marble about a central object of mass $M$. In addition, the period of revolution is also dependent upon the total mass of the fabric interior to that of the orbiting marble at a given radius, given by $\pi\sigma_0 r^2$.  When the mass of the central object is small compared to the total mass of the fabric interior to the orbiting marble ($M\ll\pi\sigma_0r^2$) Eq.~\eqref{kepler} reduces to
\be\label{vac}
T^3\propto r/\sqrt{\sigma_0}.
\ee

Notice that the two competing terms in the denominator of Eq.~\eqref{kepler} contribute equally when
\be\label{equal}
M\simeq\pi\sigma_0 r^2.
\ee
For our spandex fabric, we measured an areal mass density of $\sigma_0=0.19$ kg/m$^2$.  In our experimental analysis of the small curvature regime (discussed below), the average radius of the first revolution for a given central mass was $r\simeq 0.41$\,m.  Plugging these values into Eq.~\eqref{equal}, we find that a mass of $M\simeq 0.10$\,kg contributes equally to that of the spandex fabric.  Hence, when the mass of the central object is less than about $0.10$ kg, the mass of the spandex fabric is the dominant factor in the Kepler-like relation of Eq.~\eqref{kepler}.

In this manuscript, we experimentally analyze a range of central masses where the functional behavior of either Eq.~\eqref{white} or~\eqref{vac} is allowed to dominate.  By plotting $T^3$ vs $r_{\rm ave}^2/(M+\pi\sigma_0 r_{\rm ave}^2)^{1/2}$,  where $r_{\rm ave}$ is the average radius over a full revolution, we obtain a straight-line fit confirming the validity of Eq.~\eqref{kepler}.  The slope of that straight line yields an indirect measurement of the parameter $\alpha$ and therefore the modulus of elasticity $E$.

\subsection{The experiment in the small curvature regime.}\label{small}

For our apparatus, we removed the springs and trampoline surface from a $4$-ft mini-trampoline and secured a fabric to the frame by means of a ratchet strap.  To ensure that the fabric was given no pre-stretch, a styrofoam insert was fashioned to exactly fit the interior of the trampoline frame, forming a table-like structure.  The fabric was then draped over the frame, with the styrofoam insert in place, and fastened to the frame securely by the ratchet strap.  Once the fabric was secured, the styrofoam insert was removed and the fabric was allowed to hang freely.    It should be noted that this non-prestretched fabric sagged nearly four inches at its center, significantly warping due to its own weight.

\begin{figure}
\begin{center}
\includegraphics[width=12cm]{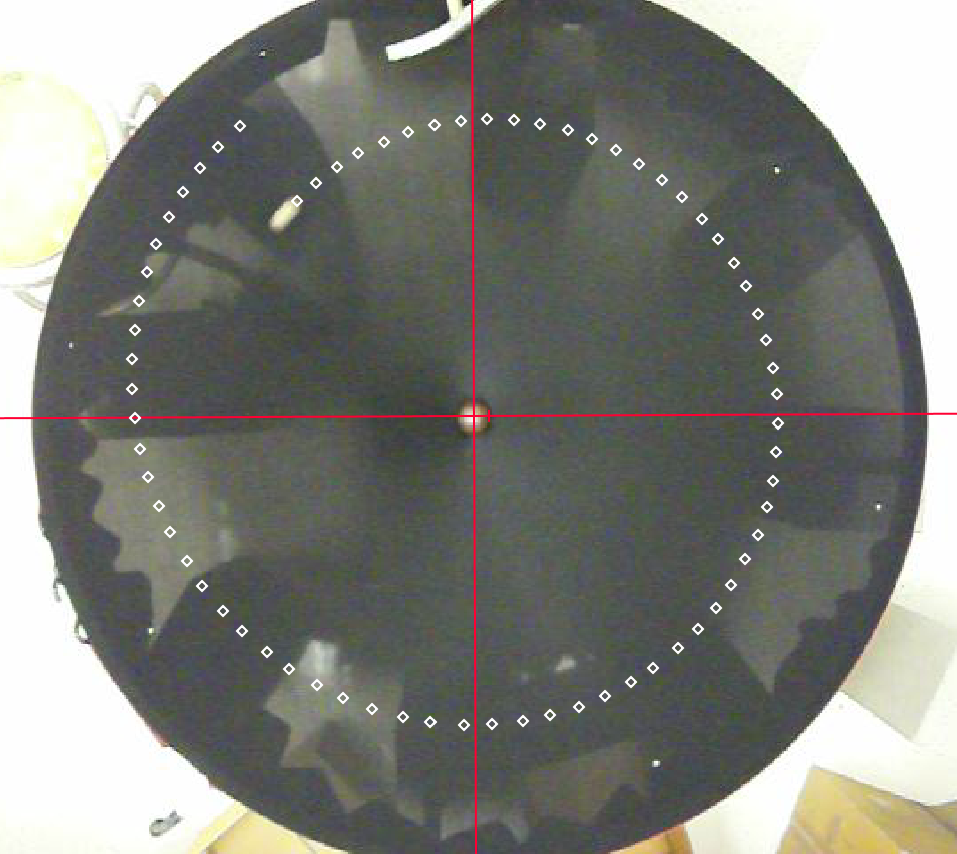}
\caption{Screenshot of a top view of a marble rolling on the spandex surface, here subjected to a central mass of $M=0.6013$\,kg.  Each diamond marks the position of the marble, in intervals of 1/30 of a second, as it orbits the central mass.  From the collection of instantaneous positions imaged here, the average radius of the orbit and the respective period of revolution can be calculated, thus yielding one data point.
\label{fig:onerev}}
\end{center}
\end{figure}

A small camera capable of taking short video clips was mounted above the fabric and leveled.   A ramp was mounted onto the perimeter of the trampoline, which allowed for a consistent set of initial conditions (speed and radius) that produced circular-like orbits.   Several short video clips of such orbits were then generated for a range of central masses, including a run with zero central mass.  The video clips were then imported into Tracker,~\cite{tracker} a video-analysis and modeling software program, where data including the particle's location as a function of time could be readily generated.  By knowing the measured diameter of the trampoline frame, the imported video could be calibrated and a proper length scale established.  As Tracker allows one to proceed through a video clip one frame at a time, the location of a small  marble could be determined as a function of time (see Fig.~\ref{fig:onerev}).   These data sets were then imported into Excel for analysis.

\begin{figure}
\begin{center}
\includegraphics[scale=1.2]{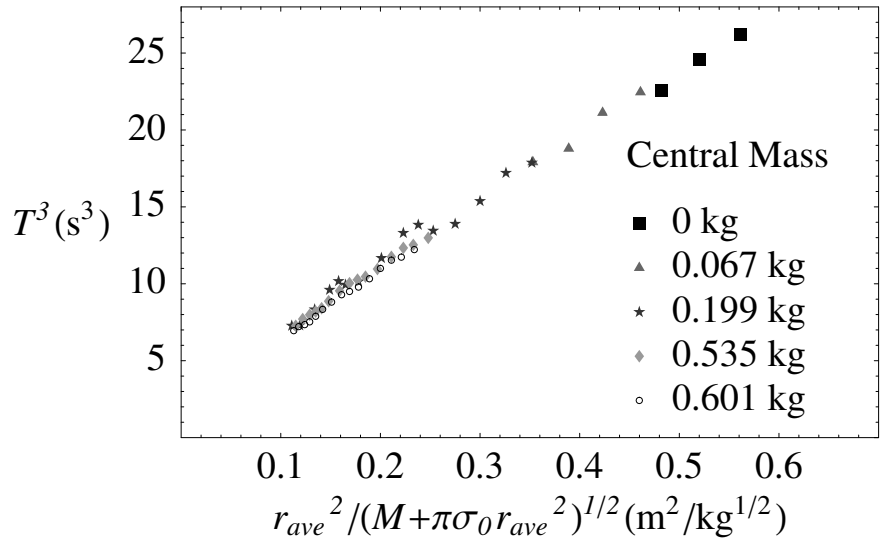}
\caption{Plot of $T^3$ vs $r_{ave}^2/(M+\pi\sigma_0 r_{ave}^2)^{1/2}$ for circular-like orbits in the small-curvature regime for central masses that range from $M=0$\,kg to $M=0.6013$\,kg.  When a straight line is fit to the data we find a slope of 45.6\,kg$^{1/2}$s$^3$/m$^2$, which corresponds to $\alpha=0.043$\,m/kg.
\label{fig:smallangle}}
\end{center}
\end{figure}

In the small curvature regime orbits with small eccentricities are challenging to obtain, so approximately 10--12 runs per central mass were imaged and the most circular-like orbit was chosen for analysis.   While Eq.~\eqref{kepler} is valid for circular orbits, in our experiments the radius of a given orbit decreases (significantly in the small curvature regime) during each revolution due to frictional losses and drag.  To account for this radial decrease and the non-zero eccentricities of the orbits, the average radius over a full revolution was calculated by averaging over the measured instantaneous radii.  The period of the respective orbit was then calculated and one data point was hence generated from these two quantities.  The initial radius for the next data point was then chosen to the radius of the previous data point after one-eighth of a revolution, and the procedure was repeated.  To remain within the small curvature regime, only $1.25$ to $2.75$ revolutions were analyzed, yielding 3--15 data points per central mass (the number of data points depended on the mass of the central object).

Employing Eq.~\eqref{kepler}, the data was plotted in a manner to yield a straight line. Figure~\ref{fig:smallangle} shows the data corresponding to circular-like orbits in the small-curvature regime for central masses that range from zero to $M=0.6013$\,kg.   When a best fit line is imposed on the data we obtain a slope of 45.6\,kg$^{1/2}$s$^3$/m$^2$, and use of Eqs.~\eqref{kepler} and~\eqref{alpha} then leads to
\bea\label{parameters}
\alpha&=&\frac{1}{2}\left(\frac{28\pi^2}{5g}\right)^3\frac{1}{\mbox{slope}^2}\simeq 0.043\,\mbox{m/kg}
\eea
and
\bea
E&=&\frac{g}{2\pi\alpha}\simeq 36\,\mbox{N/m}.
\eea

We note that we were even able to generate a circular-like orbit with no central mass, with the warping of the fabric due solely to the mass of the fabric itself.  Unfortunately, after about a revolution and a half this marble would deviate significantly from a circular-like orbit and plunge towards the center of the fabric.  Nevertheless, the first revolution and a quarter were analyzed and three data points were obtained.  Remarkably, these three data points, corresponding to zero central mass, fall on the predicted straight line  (see Fig.~\ref{fig:smallangle}).

\section{Circular orbits in the large curvature regime}\label{orbits2}

We now explore the behavior of Eq.~\eqref{shape} in the large curvature regime of the fabric where the warping is large, namely, when $z'(r)\gg 1$.  In this large $z'$ regime we can drop the aerial mass density term as it can be shown to be insignificant in this large curvature regime.  Then we expand the square root in in powers of $1/z'$ and keep the lowest order terms to obtain
\be\label{largeslope}
z'(r)\simeq \frac{M\alpha}{r}+1.
\ee 
Plugging this expression into Eq.~\eqref{cirEoM} and rearranging, we obtain a Kepler-like relation quite different to that of Eq.~\eqref{kepler}.  In the large curvature regime we now have
\be\label{kepler2}
T=\left(\frac{28\pi^2}{5g}\right)^{1/2}\frac{r}{\left(M\alpha+ r\right)^{1/2}}.
\ee
Similar to the case of the small curvature regime, this Kepler-like relation has two competing terms in the denominator of the right hand side.  Notice that for $r\ll M\alpha$, when the marble is deep within the well, Eq.~\eqref{kepler2} reduces to
\be\label{laplace}
T\propto r/\sqrt{M}.
\ee
As previously discussed,\cite{GW, DL} this result corresponds to the solution of the 2D Laplace equation with cylindrical symmetry. It is interesting to note that Eq.~\eqref{kepler2} is \textit{not} valid when $r\gg M\alpha$ because when $M\alpha/r\ll 1$, Eq.~\eqref{largeslope} reduces to $z'(r)\simeq 1$, whereas this equation was derived by assuming $z'\gg 1$.

We now wish to explore circular orbits in this large curvature regime for a range of large central masses where $M\alpha\simeq r$.  Notice that in order to generate a plot $T$ vs $r_{\rm ave}/(M\alpha+ r_{\rm ave})^{1/2}$ from the data, one needs the value of the parameter $\alpha$.  If the modulus of elasticity is constant for a fabric, then the value of $\alpha$ obtained experimentally in the small curvature regime can be used.  This, however, turns out not to be the case.

In Sec.~\ref{Shape} we arrived at a differential equation whose solution determines the shape of the fabric.  Equation~\eqref{shape} was obtained by considering the elastic and gravitational potential energies of an elastic fabric and the gravitational potential energy of the central mass placed on its surface.  In arriving at Eq.~\eqref{total} we assumed that the modulus of elasticity $E$ (and therefore $\alpha$) was constant for a given elastic surface.  Below we question this assumption and determine the value of $\alpha$ experimentally.  By directly measuring the stretch of the fabric for a wide range of central masses, we obtain the value of $\alpha$ graphically and find that the modulus of elasticity varies and is a function of the stretch.

\subsection{A direct measurement of the modulus of elasticity}

\begin{figure}
\begin{center}
\centering
\includegraphics[width=10cm]{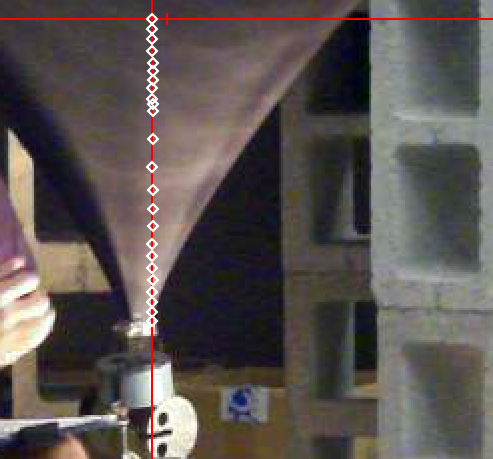}
\caption{Screenshot of a side view of the spandex surface, imaged here while subjected to a central mass of $M=7.774$\,kg.  Each diamond marks the location of a point (upper right corner of adjustable pipe clamp) on the spandex fabric when subjected to different central masses.  The top 10 points range from $M_0=0.274$\,kg to $M=1.174$\,kg in $0.1$-kg intervals; the bottom 14 points range from $M=1.274$\,kg to $M=7.774$\,kg in $0.5$-kg intervals.
\label{fig:z(M)image}}
\end{center}
\end{figure}

\begin{figure}
\begin{center}
\centering
\includegraphics[scale=1.2]{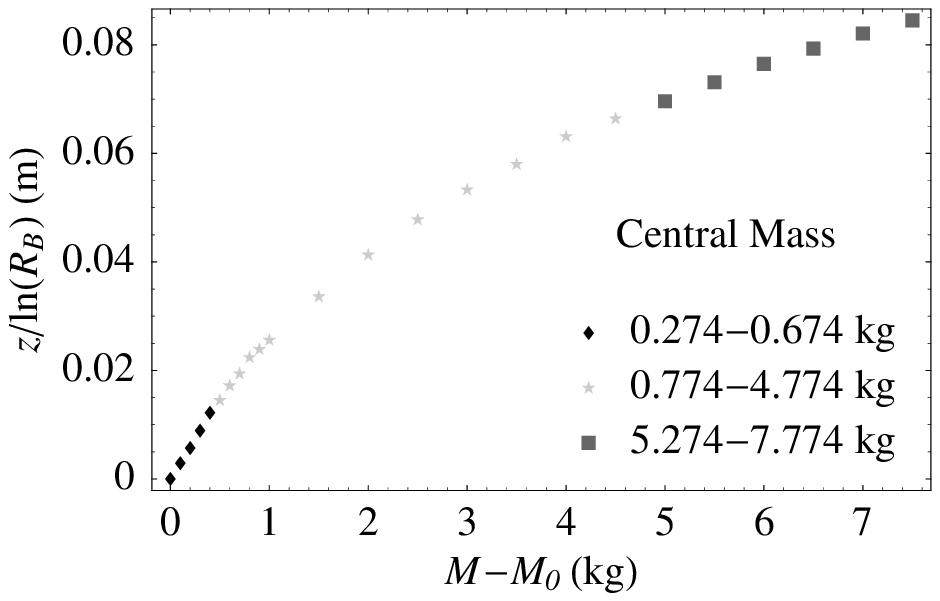}
\caption{Plot of $z(M)/\ln(R_B)$ versus $(M-M_0)$ for central masses ranging from $M=M_0=0.274$\,kg to $M=7.774$\,kg.  The local slope of this curve yields the value of $\alpha$ and therefore the modulus of elasticity ($\alpha\propto 1/E$).  As is clear from the data, the value of the modulus of elasticity of the spandex fabric is \textit{not} constant and varies with the stretch of the fabric.
\label{fig:z(M)}}
\end{center}
\end{figure}

Equation~\eqref{largeslope} can be integrated to yield the relation
\be\label{z(r)}
z(r)=M\alpha\ln(r)+r+c,
\ee
where $c$ is a constant of integration whose value is determined from a relevant boundary condition.  Equation~\eqref{z(r)} yields the height of the fabric as a function of radius for a given mass in the large curvature regime.  If one specifies the boundary condition $z(R)=0$, where $R$ is radius of the suspended fabric, then Eq.~\eqref{z(r)} takes the form of the shape function presented in Ref.~\citenum{DL} in the large $z'$ limit of Eq.~\eqref{shape}.

Instead of using the boundary condition $z(R)=0$, we consider Eq.~\eqref{z(r)} as giving the height of the fabric as a function of the central mass $z(M)$, for a given radius of the fabric.  Using a small (fixed) radius $R_B$ and taking $z(M_0)=0$, where $M_0$ is the smallest central mass probed, we find that
\be\label{z(M)}
\frac{z(M)}{\ln(R_B)}=(M-M_0)\alpha.
\ee
Given that $R_B$ is a measured constant, the left-hand-side of this expression is simply the scaled height of a point on the fabric.  Thus, if the value of the parameter $\alpha$ is constant then plotting $z(M)/\ln(R_B)$ versus $(M-M_0)$ should yield a straight line.

To determine $\alpha$, we oriented our camera to obtain a side view of the fabric's surface and obtained images of the surface using a wide range of central masses.  Figure~\ref{fig:z(M)image} shows a screenshot of the warped fabric with a central mass of $M=7.774$\,kg.  These images are then imported into Tracker to obtain the heights of the fabric at a fixed radius ($R_B$) as a function of the central mass $M$.  Figure~\ref{fig:z(M)} shows the data for $R_B=0.0196$\,m and central masses ranging from $M_0=0.274$\,kg to $M=7.774$\,kg.  As is clear from the data, the slope of the curve---and therefore the value of $\alpha$ and the modulus of elasticity---is \textit{not} constant for the range of masses considered.

If we restrict our attention to a small range of masses the value of $\alpha$ is seen to be relatively constant.  For example, in the range $M_0=0.274$\,kg to $M=0.674$\,kg, a best-fit line gives a value of $\alpha=0.030$\,m/kg in the small-central-mass regime.  This direct measurement of $\alpha$ is only about 2/3 of the value (0.043\,m/kg) found earlier.   Because the slope of the curve in Fig.~\ref{fig:z(M)} decreases with increasing central mass, we would expect that the value of $\alpha$ obtained from the slope of the best-fit line would increase slightly if smaller central masses were used.  In the mass range $M=5.274$\,kg to $M=7.774$\,kg, corresponding to the range of central masses to be explored in the large-curvature regime, the slope is again seen to be fairly constant with a value of $0.006$\,m/kg.

\subsection{The experiment in the large curvature regime}

The experimental setup and procedure for probing the large curvature regime of the spandex fabric was largely the same as it was for probing the small curvature regime, with a few exceptions.  Here we probe very small radii for a selection of large central masses. To accomplish this, a small spherical central object of radius $\sim 1.8$\,cm was positioned in the center of the fabric.  This object helped maintain the desired shape of the circular well.  A sling, capable of supporting several kilograms, was fashioned and then fastened to the central object from below the fabric by means of an adjustable pipe clamp\cite{Gary}  (the adjustable clamp was lined with rubber to allow for a snug fit and to prevent tearing of the fabric).   We note that the total mass of the small central object, the sling, and the adjustable clamp was $M_0=0.274$\,kg and yielded a lower mass bound when measuring the modulus of elasticity.

Unlike in the small-curvature regime, circular orbits with small eccentricities are easy to obtain in the large curvature regime.  Additionally, the radius of a given orbit does not decrease substantially with each revolution, which allows us to obtain data for numerous circular-like orbits per run.  As was done in the small curvature regime, the average radius over a full revolution was calculated by averaging over the measured instantaneous radii and the period of the respective revolution was obtained from the video data.  The initial radius for the sequential data point was chosen to be one-half of a revolution ahead (compared to one-eighth of a revolution in the small curvature regime) of the previous initial radius and the aforementioned procedure was then repeated.

\begin{figure}
\begin{center}
\includegraphics[scale=1.2]{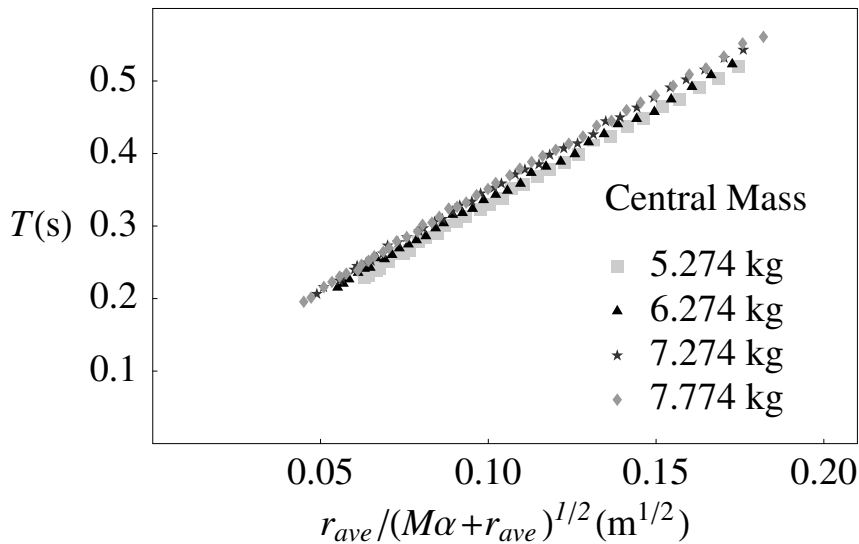}
\caption{Plot of $T$ versus $r_{ave}/(M\alpha+r_{ave})^{1/2}$ for circular-like orbits in the large-curvature regime for central masses that range from $M=5.274$\,kg to $M=7.774$\,kg.  When $\alpha=0.006$\,m/kg we find a straight line with slope 2.62\,s/m$^{1/2}$, which deviates from the theoretical value by $\sim 10\%$.
\label{fig:largeangle2}}
\end{center}
\end{figure}

Employing Eq.~\eqref{kepler2}, the data was plotted in a manner to yield the predicted straight line.  Figure~\ref{fig:largeangle2} shows the data corresponding to circular-like orbits in the large-curvature regime for central masses ranging from $M=5.274$ kg to $M=7.774$ kg.  (As previously discussed, the value of $\alpha$ was determined graphically from the slope of Fig.~\ref{fig:z(M)} in the large-mass regime to be $\alpha=0.006$ m/kg.)  A best-fit line applied to these data yields a slope of $2.62$\,s/m$^{1/2}$, which deviated from the theoretical value of $(28\pi^2/5g)^{1/2}=2.37$\,s/m$^{1/2}$ by about 10\%.  Although the agreement is not perfect, we note that if Eq.~\eqref{parameters} is naively used for $\alpha$ in the large curvature regime, we find a slope of $4.59$\,s/m$^{1/2}$, which corresponds to a $94\%$ error from the theoretical value.

\section{conclusion}

The main findings of the work are two-fold.  First, when analyzing circular orbits of a marble on an elastic fabric in the small curvature regime, the contribution of the mass of the elastic fabric interior to the orbiting marble is relevant to the analysis and can dominate over the contribution of the central mass.  Second, we found that the modulus of elasticity for a spandex fabric is \textit{not} constant and is itself a function of the stretch.  In analyzing the small and large curvature regimes we found that the value of the modulus of elasticity was approximately constant in each regime.  Therefore, the above analysis is valid when applied to each regime separately.   We note that it would be interesting to generalize the analysis by allowing for a variable modulus of elasticity, and therefore a variable parameter of the form $\alpha(z')$.

Although orbits on any 2D cylindrically-symmetric surface have been shown to deviate from the Keplerian orbits of planetary motion for the general case of noncircular orbits,\cite{English} the analogy between the two remains conceptually useful for the beginning student of general relativity.  Apart from GR, the mathematical evaluation of the shape of the elastic fabric and the motion of a marble rolling on this surface offer two rich examples of the calculus of variations and Lagrangian dynamics.  As such, the work presented here could be included in an advanced dynamics course for undergraduates.

\appendix*

\section{Circular orbits in general relativity}\label{AdS}

In this section, we briefly outline the general relativistic treatment for obtaining circular orbits about a spherically-symmetric massive object, such as a non-rotating star or black hole, in the presence of a constant vacuum energy density (a cosmological constant).  Our aim here is to merely introduce the reader to the essence of particle orbits in general relativity (GR), and to outline the conditions necessary for obtaining circular orbits.  We arrive at an exact Kepler-like expression for the orbiting body that can then be compared to the Kepler-like relation of the marble on the warped elastic fabric in the small curvature regime.

Historically speaking, Einstein first introduced a cosmological constant into the field equations of GR in order to obtain a static cosmological model to describe the seemingly static universe of his time.  This notion of a cosmological constant was later abandoned by Einstein, and famously denounced as the greatest blunder of his life, after the discovery of the expansion of the universe.  After the later discovery that the expansion of the universe is in fact accelerating, the cosmological constant has re-emerged into the Einstein field equations, only now incorporated into GR as a vacuum energy density.   This relabeling of the cosmological constant arises from the prediction of zero-point fluctuations in quantum field theory, which give rise to a vacuum energy density.  Mathematically speaking, the cosmological constant and a constant vacuum energy density are proportional to one another and related by the expression
\be
\rho_{vac}=\frac{\Lambda}{8\pi G},
\ee
where $G$ is Newton's universal constant and we are working in units where $c=1$.

In general relativity gravity is described as the warping of space and time due to the presence of matter and energy.  To understand particle orbits about a central object, one must first arrive at the metric describing that particular spacetime.   The metric represents an infinitesimal ``distance'' in a four-dimensional sense and completely describes the spacetime geometry.  The metric exterior to a spherically-symmetric massive object, in the presence of a constant vacuum energy, is of the form
\be\label{metric}
ds^2=-\left(1-\frac{2GM}{r}-\frac{\Lambda}{3}r^2\right)\;dt^2+\left(1-\frac{2GM}{r}-\frac{\Lambda}{3}r^2\right)^{-1}dr^2+r^2\left(d\theta^2+\sin^2\theta\;d\phi^2\right),
\ee
where $M$ is the mass of the central object and $\Lambda$ the cosmological constant.  We are working in the spherical-polar coordinates $(t,r,\theta,\phi)$.  

Notice that $\Lambda$ can be either positive or negative.  For $\Lambda>0$ the above metric represents a Schwarzschild-de Sitter space-time, and for $\Lambda<0$ the metric represents a Schwarzschild-anti-de Sitter (AdS) space-time.\cite{Stuchlik, Hackmann, Cruz}  Observational evidence suggests that we live in a universe with a nonzero, positive vacuum energy density corresponding to $\Lambda>0$.  This positive vacuum energy density gives rise to a repulsive `force' in the late universe and is needed to account for the accelerated expansion within the framework of GR.   In this manuscript, however, we will be more interested in the Schwarzschild-AdS solution as the Kepler-like relation for this space-time is somewhat analogous to the Kepler-like relation for the marble on the fabric in the small curvature regime.

By normalizing the four-velocity and employing conservation of energy and angular momentum, which arise from the fact that the metric is independent of the time $t$ and angular coordinate $\phi$,\cite{hartle, carroll} one arrives at a radial equation of motion of the form
\be\label{energy}
\mathcal{E}=\frac{1}{2}\left(\frac{dr}{d\tau}\right)^{2}+V_{\rm eff}(r),
\ee
where
\be
V_{\rm eff}(r)=-\frac{GM}{r}+\frac{\ell^2}{2r^2}-\frac{GM\ell^2}{r^3}-\frac{\Lambda}{6}(\ell^2+r^2).
\ee
Here $\tau$ is the proper time and $\mathcal{E}$ is a constant of the motion.  In arriving at Eq.~\eqref{energy}, we chose the motion to occur in the equatorial plane, dictated by $\theta=\pi/2$.   It should be noted that Eq.~\eqref{energy} is an exact expression and is effectively a statement of conservation of mechanical energy per unit mass for a classical particle moving in one dimension.

Notice that by setting $\Lambda=0$, the effective potential energy per unit mass $V_{\rm eff}(r)$ contains an additional $1/r^3$ contribution to the regular $1/r$ and $1/r^2$ terms found in a Newtonian treatment of particle orbits.  This additional GR term offers a small $r$ modification to non-relativistic Newtonian orbits.   For a non-zero cosmological constant, the last term in the effective potential energy offers a large $r$ modification to the Schwarzschild orbits of a body about a non-rotating, spherically-symmetric central mass.

To arrive at circular orbits, one sets $V_{\rm eff}'(r)=0$ and $\mathcal{E}=V_{\rm eff}(r)$.\cite{Schwarzschild} Employing these conditions and incorporating conservation of energy and angular momentum, one arrives at a Kepler-like relation of the form
\be\label{AdSkepler}
T^2\propto\frac{r^3}{\left(M-\Lambda r^3/3G\right)},
\ee 
where $T$ is the period of revolution measured by a stationary observer (see Ref.~\onlinecite{Cruz} for details on arriving at the above expression).  It should be noted that this expression reduces to Kepler's third law when $\Lambda$ is set equal to zero.\cite{kepler}  Not surprisingly, Eq.~\ref{AdSkepler} shows that the period $T$ depends on the radius of the orbit $r$ and the mass $M$ of the central object.   In addition, the period also depends on the vacuum energy, whose functional dependence is given by the $\Lambda r^3$ term in the denominator.  Naively, one can identify the $\Lambda r^3/3G$ term as twice the total mass of the vacuum interior to that of the orbiting body at a given radius.  However, there is a subtlety in calculating volume in general relativity where one must account for  the non-Euclidean nature of space-time.   A similar subtlety occurs when calculating the interior mass of a non-rotating spherically-symmetric star, with the discrepancy arising from the binding energy of the star itself.\cite{carroll}  Notice that for $\Lambda<0$ the period of revolution is smaller for a given mass $M$ and radial distance $r$ as compared to a space-time with $\Lambda=0$.  This is consistent with the fact that a negative cosmological constant gives rise to a greater attractive ``force'' on the orbiting body.

Recall the functional form of the Kepler-like expression for a small marble orbiting on the warped fabric in the small curvature regime is of the form
\be\label{kepler.}
T^3\propto\frac{r^2}{\left(M+\pi \sigma_0 r^2\right)^{1/2}}.
\ee
Upon comparison of this expression with Eq.~\eqref{AdSkepler}, we find that the areal mass density $\sigma_0$ of the fabric plays the role of a negative cosmological constant of the Schwarzschild-AdS space-time when the conceptual analogy between particle orbits in GR and rolling marble orbits on a warped fabric is employed.  This makes sense as a negative cosmological constant gives rise to an attractive ``force'' on the orbiting particle.  We note, however, that the functional dependence connecting the period of revolution to the orbital radius and the mass of the central object for the rolling marble differs dramatically from that of an orbiting body around a Schwarzschild-AdS space-time.  We further note that Eq.~\eqref{AdSkepler} represents an exact expression, whereas Eq.~\eqref{kepler.} amounts to an approximate relation in the small-curvature regime.


\begin{thebibliography}{19}
\bibitem{noprestretch}
Notice that the analysis presented in Ref.~\onlinecite{DL} allowed for the spandex fabric to be pre-stretched, which dramatically alters the shape of the fabric.  In the present work we ensured that the spandex fabric was given zero pre-stretch.
\bibitem{Schwarzschild}
As is the case for the Schwarzschild metric, $r<3GM$ yields no circular orbits.
\bibitem{kepler}
It may be surprising for the reader to learn that Kepler's third law is of the same form in general relativity as it is in Newtonian gravitation.
\end{thebibliography}
\end{document}